\documentclass[aps,prl,amsmath,amsfonts,superscriptaddress,twocolumn]{revtex4}
\usepackage{epsfig,psfig,graphicx}

\newcommand{\beq}{\begin{equation}}
\newcommand{\enq}{\end{equation}}

\begin{document}

\title{Atom-molecule theory of broad Feshbach resonances}

\author{G.M. Falco}

\affiliation{Institute for Theoretical Physics, Utrecht
University, Leuvenlaan 4, 3584 CE Utrecht, The Netherlands}

\author{H.T.C. Stoof}

\affiliation{Institute for Theoretical Physics, Utrecht
University, Leuvenlaan 4, 3584 CE Utrecht, The Netherlands}

\begin{abstract}
We derive the atom-molecule theory for an atomic gas near a broad
Feshbach resonance, where the energy dependence of the
atom-molecule coupling becomes crucial for understanding
experimental results. We show how our many-body theory
incorporates the two-atom physics exactly. In particular, we
calculate the magnetic moment of a two-component gas of $^{6}$Li atoms for a
wide range of magnetic fields near the broad Feshbach resonance at
about $834$ Gauss. We find excellent agreement with the experiment
of Jochim {\it et al.} [Phys.~Rev.~Lett. {\bf 91}, 240402 (2003)].
\end{abstract}


\maketitle

{\it Introduction.} --- By sweeping an external magnetic field in
the right direction across a Feshbach resonance, it is possible to
create ultracold diatomic molecules in an atomic gas
\cite{greiner2003,fermi1,fermi2,fermi3,zwierlein2003}. The reason for this formation
of molecules is that, by changing the magnetic field, the energy
of the molecular state that causes the Feshbach resonance can be
moved from above to below the threshold of the two-atom continuum
\cite{stwalley1976,tiesinga1993}. In an atomic Fermi gas this
experimental control over the location of the molecular energy
level offers the exciting possibility to study in detail the
crossover between the Bose-Einstein condensation (BEC) of diatomic
molecules and the Bose-Einstein condensation of atomic Cooper
pairs, i.e., the Bardeen-Cooper-Schrieffer (BCS) transition
\cite{stoof1996,timmermans2001,ohashi2002,milstein2002}. In
particular, the very broad resonances of $^{6}$Li at about $834$
Gauss is for this purpose used by a number of experimental groups
\cite{ketterle2004,grimm2004,thomas2004,salomon2004,hulet2004}.

The use of this broad Feshbach resonance, however, seriously
complicates the theoretical analysis of the experimental results,
because for such a broad resonance the atom-molecule coupling can
no longer be assumed to be independent of the relative
energy of the colliding atoms. Physically, this implies that the
atom-molecule coupling shows retardation effects in this case.
Dealing with these retardation effects requires the development of
an effective atom-molecule theory that is suitable for many-body
calculations and at the same time incorporates the exact
energy-dependence of the two-body scattering process. It is the
main purpose of this Letter to show how we can arrive at such an
effective atom-molecule theory by generalizing the quantum field
theory developed previously for atomic Bose gases in
Refs.~\cite{duine2003a,duine2003b} and extended to atomic Fermi
gases in Refs.~\cite{falco2004a,falco2004b}.

To achieve this goal we need to determine the energy and lifetime
of the Feshbach molecule in the presence of the ultracold atomic
gas, i.e., we need to determine the molecular selfenergy. We,
therefore, first present a many-body approach that can be used to
calculate the selfenergy exactly. Having the above-mentioned
application in mind, we focus here on a two-component Fermi gas.
However, our formalism applies also to an atomic Bose gas or to an
atomic Bose-Fermi mixture. After we have obtained the exact expression
for the molecular selfenergy in the presence of a medium, we then
consider the two-body limit of the theory. The reason for
considering this limit is twofold. First, any accurate many-body
theory for the BEC-BCS crossover in atomic $^{6}$Li must
incorporate the two-body physics exactly to be able to make a
successful comparison with experiments. Second, in this limit we
are already able to compare the theory with experimental
data on the magnetic moment of an atomic $^{6}$Li gas and with coupled-channels
calculations of the molecular binding energy. 
We find excellent agreement in both cases. We, therefore, conclude that
our quantum field theory gives an accurate account of the
retardation effects occurring near a broad Feshbach resonance. 
The theory can thus also be used to investigate the BEC-BCS crossover
phenomenon for a broad Feshbach resonance, where subtle two-body
threshold effects and strong-coupling
many-body physics merge together.

{\it Molecular selfenergy.} --- From now on we consider an
incoherent mixture of fermionic atoms in two different hyperfine
states, which we denote by $|\uparrow\rangle$ and
$|\downarrow\rangle$. If this two-component Fermi gas is in the
vicinity of a Feshbach resonance, the effective quantum field
theory of the gas is defined by means of the atom-molecule
Hamiltonian \cite{duine2003b}
\begin{eqnarray}
\label{eq:Ham} H&=&\sum_{\mathbf{k},\sigma}
\left(\epsilon_{\mathbf{k}}-\mu\right)
a^{\dagger}_{\mathbf{k},\sigma} a_{\mathbf{k},\sigma}\nonumber\\
&+&\frac{1} {V}\sum_{\mathbf{k},\mathbf{k'},\mathbf{q}}
V_{\rm{bg}}\left(\mathbf{q}\right)
a^{\dagger}_{\mathbf{k+q},\uparrow}
a^{\dagger}_{\mathbf{k'-q},\downarrow}
a_{\mathbf{k'},\downarrow}a_{\mathbf{k},\uparrow}
\nonumber\\
&+&\sum_{\mathbf{k}}\left(\frac{\epsilon_{\mathbf{k}}}{2}+\epsilon_{\rm{bare}}
+\Delta\mu B-2\mu\right) b^{\dagger}_{\mathbf{k}} b_{\mathbf{k}}
\nonumber\\
&+&\frac{g_{\rm bare}}{\sqrt{V}}\sum_{\mathbf{k},\mathbf{q}}
(b^{\dagger}_{\mathbf{k}} a_{\frac{\mathbf
k}{2}+\mathbf{q},\uparrow}a_{\frac{\mathbf
k}{2}-\mathbf{q},\downarrow}+{\rm h.c.}) .
\end{eqnarray}
Here, $a^{\dagger}_{\mathbf{k},\sigma}$ is the creation operator
of an atom with momentum $\hbar\mathbf{k}$ and in the hyperfine
state $|\sigma\rangle$, and $b^{\dagger}_{\mathbf{k}}$ is the
creation operator of a bare or Feshbach molecule. In addition,
$\epsilon_{\mathbf{k}}$ is the kinetic energy of an atom, $\mu$ is
the chemical potential, $V_{\rm{bg}}$ is the nonresonant or
background interaction between the atoms, $g_{\rm bare}$ is the
bare atom-molecule coupling, $B$ is the external magnetic field,
$\epsilon_{\rm{bare}}$ is the energy of a bare molecule with zero total
momentum, and $\Delta \mu$ is the difference in magnetic moment
between the bare molecule and two atoms.

\begin{figure}[h]
\epsfig{figure=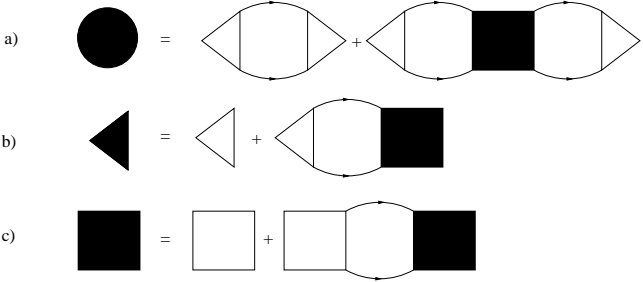,width=8.5cm} \caption{Diagrammatic
representation of (a) the exact molecular selfenergy,
 (b) the renormalization of the atom-molecule coupling, and
(c) the Bethe-Salpeter equation for the one-particle irreducible vertex function.
Note that 
triangles denote the atom-molecule coupling, that squares refer to the two-particle vertices,
and that the internal lines connecting these quantities represent fully dressed atomic propagators.
\label{diagram}}
\end{figure}

Given the above atom-molecule Hamiltonian, the molecular
selfenergy can, in the normal state of the atomic gas, be
calculated from the following procedure. We consider here only the
normal state of the atomic gas, because we ultimately want to
compare our results with the experiment of Jochim {\it et al.}
\cite{fermi3}. The generalization to the superfluid state is also
of interest, but will be reported elsewhere. We start by
expressing the selfenergy $\hbar \Sigma_{\rm{m}}$ of a single molecule in
the atomic gas in terms of the exact atomic Green's functions
$G_{\sigma}$ and the exact (four-point) vertex function
$\Gamma_{\sigma,\sigma'}$ as
\begin{eqnarray}
\label{eq:exact0} \hbar\Sigma_{\rm{m}}= - \frac{1}{\hbar} g_{\rm {bare}}
\left( G_{\uparrow} G_{\downarrow} - \frac{1}{\hbar} G_{\uparrow}
G_{\downarrow} \Gamma_{\uparrow,\downarrow} G_{\uparrow}
G_{\downarrow} \right) g_{\rm {bare}}~.
\end{eqnarray}
Note that we are using an operator notation here that suppresses
the dependence on momenta and Matsubara frequencies of the various
quantities involved. The diagrammatic equivalent of
Eq.~(\ref{eq:exact0}) is shown in Fig.~\ref{diagram}a.

This result
can be clarified by introducing the exact atom-molecule coupling
$g$ by means of (see Fig.~\ref{diagram}b)
\begin{equation}
g=\left(1-\frac{1}{\hbar}\Gamma_{\uparrow,\downarrow} G_{\uparrow}
G_{\downarrow}\right)g_{\rm bare}~,
\end{equation}
since then the selfenergy can be put into the expected form
\begin{equation}
\hbar\Sigma_{\rm{m}} =- \frac{1}{\hbar} g_{\rm bare} G_{\uparrow}
G_{\downarrow} g~.
\end{equation}
Because the exact vertex function is one-particle irreducible, it
satisfies the Bethe-Salpeter equation
\begin{eqnarray}
\label{eq:BetheSalpeter} \Gamma_{\uparrow,\downarrow} =
\Gamma^{\rm ir}_{\uparrow,\downarrow} - \frac{1}{\hbar}
\Gamma^{\rm ir}_{\uparrow,\downarrow} G_{\uparrow} G_{\downarrow}
\Gamma_{\uparrow,\downarrow}~,
\end{eqnarray}
whose diagrammatic equivalent is shown in
Fig.~\ref{diagram}c. The quantity $\Gamma^{\rm
ir}_{\sigma,\sigma'}$ is the exact vertex function that is
two-particle irreducible in the particle-particle channel. From
Eqs.~(4) and (5) it appears that the selfenergy depends on the
bare atom-molecule vertex. However, the definition of the exact
atom-molecule coupling can be used to eliminate the bare coupling
from the theory. 

{\it Atomic physics.} --- 
The two-body limit is obtained by replacing in the above procedure
the exact atomic Green's functions by the noninteracting atomic
Green's functions, and the two-particle irreducible vertex
function by $V_{\rm bg}$. Performing the required calculations, we
then find that the energy-dependent dressed atom-molecule coupling
becomes
\begin{eqnarray}
\label{eq:gexp} g^{\rm{2B}}\left(E\right)=g(B) \frac{1}{1-\left[a_{\rm
bg}(B)/\hbar\right]\sqrt{-mE^+}}~,
\end{eqnarray}
where $E^+ = E + i0$, and both the background scattering length
$a_{\rm bg}(B)$ of the background interaction $V_{\rm bg}$ and the
zero-energy coupling constant $g(B)$ are known experimental
quantities. Moreover, the molecular selfenergy obeys
\begin{eqnarray}
\label{eq:self}
\hbar\Sigma^{\rm{2B}}\left(E\right)-\hbar\Sigma^{\rm{2B}}\left(0\right)=\frac{\eta
(B)\sqrt{-E^+}}{1+|a_{\rm bg}(B)/\hbar|\sqrt{-mE^+}}~,
\end{eqnarray}
where the energy $\eta^2(B) =g^4(B) m^3/16\pi^2\hbar^{6}$ defines an
important energy scale in the problem, which is related to the
width of the Feshbach resonance.

The energy of the dressed molecular state with zero kinetic energy
$\epsilon_{\rm {m}}$ is determined by the poles of the full
molecular propagator, which is equivalent to solving the equation
\begin{eqnarray}
\label{eq:root} \epsilon_{\rm{m}} = \delta(B)+
\frac{\eta(B)\sqrt{-\epsilon_{\rm m}^+}}{1+|a_{\rm bg}(B)/\hbar|\sqrt{-m\epsilon_{\rm m}^+
}},
\end{eqnarray}
where $\delta(B)= \Delta\mu \,B +
\epsilon_{\rm{bare}}+\hbar\Sigma^{\rm{2B}}\left(0\right) \equiv
\Delta\mu\left(B-B_0\right)$ is known as the detuning and $B_0$ is
the magnetic field location of the Feshbach resonance. For
positive detuning there only exists a solution with a negative
imaginary part, in agreement with the fact that the molecule
decays when its energy is above the two-atom continuum threshold.
For negative detuning the dressed molecular propagator has a real
pole at negative energy corresponding to the bound-state energy of the
dressed molecule. 

Near resonance the bound-state energy
$\epsilon_{\rm m}(B)$ becomes small and we are allowed to put the
right-hand side of Eq.~(\ref{eq:root}) equal to zero. In this
approximation the location of the pole can be found analitically and yields
the expected result
\begin{eqnarray}
\label{eq:bind} \epsilon_{\rm m}(B)=-\frac{\hbar^2}{m a^2(B)}~,
\end{eqnarray}
where the total scattering length $a(B)$ is given by
\begin{eqnarray}
\label{eq:sclen}
\frac{4\pi\hbar^{2}a(B)}{m}=\frac{4\pi\hbar^{2}a_{\rm {bg}}(B)}{m}-\frac{g^2(B)}{\delta(B)}~.
\end{eqnarray}
The general solution, which is required when we want to know the
bound-state energy over a large range of magnetic fields, can only
be found numerically as we show shortly. The residue of the pole
is in general given by
\begin{eqnarray}
\label{eq:zeta}
Z(B)=\left[1-\frac{\partial\hbar\Sigma^{\rm{2B}}\left(E\right)
}{\partial E} \right]_{E=\epsilon_{\rm_m}(B)}^{-1}
\end{eqnarray}
and it is always smaller than one. This is because the dressed
molecular state near the Feshbach resonance obeys
\begin{multline}
\label{eq:wavefctmol}
\langle {\bf r}| \chi_{\rm m}; {\rm dressed} \rangle\simeq
                 \sqrt{Z(B)} \chi_{\rm m}({\bf r}) |{\rm closed} \rangle
                 \nonumber \\
 + \sqrt{1-Z(B)}\frac{1}{\sqrt{2 \pi a(B)}}
\frac{e^{-r/a(B)}}{r}
                           |{\rm open} \rangle ~,
\end{multline}
where $\chi_{\rm m}({\bf r})$ denotes the wavefunction of the bare
molecular state in the closed channel of the Feshbach problem. The
dressed molecular state therefore only contains with an amplitude
$\sqrt{Z(B)}$ the bare molecular state $|\chi_{\rm m};{\rm closed}
\rangle$ of the closed channel.

{\it Magnetic moment.} --- 
Since the open and the closed channels of the Feshbach resonance in
atomic $^6$Li experiments correspond in an excellent approximation
to the electronic triplet and singlet states, respectively, we
have $\Delta \mu= 2\mu_B$, with $\mu_B$ the Bohr magneton. As a
result, the magnetic moment of the dressed molecules $\mu_{\rm m}$
as a function of magnetic field is equal to
\begin{eqnarray}
\label{eq:mu} \mu_{\rm m}(B)=2 \mu_{B}(1-Z(B)).
\end{eqnarray}
Close to resonance where $Z(B) \ll 1$ and the contribution of the
open channel becomes large, the magnetic moment approaches the
value $2 \mu_B$ characteristic of the triplet state. Far off
resonance we have $Z(B) \simeq 1$ and the dressed state is almost
equal to the bound state of the closed channel potential whose
electronic spin is a singlet. In that case we thus have that
$\mu_{\rm m}(B)$ goes to zero.

The magnetic moment of the molecules obtained with the very broad
Feshbach resonance at about $834$ Gauss in a gas of $^{6}$Li atoms
has been measured by Jochim {\it et al.} \cite{fermi3}. For such
a broad resonance, the background scattering length $a_{\rm{bg}}(B)$
and the zero-energy atom-molecule coupling constant $g(B)$, which, together
with the magnetic field $B_0$, constitute the only input
parameters of our theory, depend strongly on the magnetic field
over the relevant experimental range. In contrast to narrow
resonances, neglecting these magnetic field dependences is not
sufficiently accurate here. Therefore, we use the field-dependence
of $a_{\rm{bg}}(B)$ that was calculated by Marcelis {\it et
al.} from a careful analysis of the experimental knowledge of
the singlet and triplet interatomic potentials \cite{Marcelis2004}.
The field-dependent
atom-molecule coupling constant is deduced from $a_{\rm{bg}}(B)$ by using
Eq. (\ref{eq:sclen}) and the known total scattering length $a(B)$ \cite{randy}.

\begin{figure}[h]
\epsfig{figure=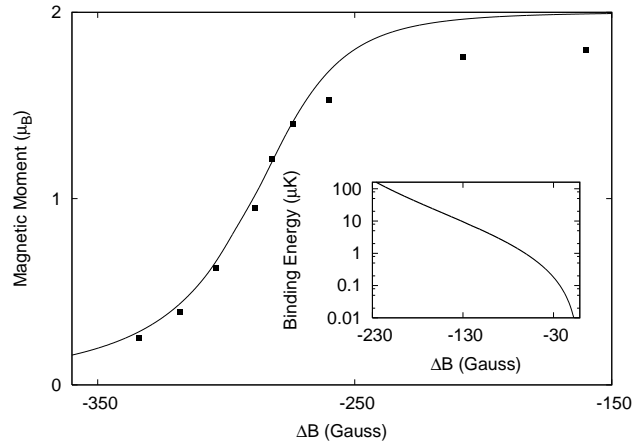,width=8.5cm} \caption{\rm Magnetic
moment of the dressed molecules as a function of the external magnetic
field $\Delta B=(B-B_0)$. The dots represent the experimental
data of Ref.~\cite{fermi3}, while the solid line is our theoretical result. For the field-dependent 
background scattering length $a_{\rm{bg}}(B)$ we have used
the parametric expression
obtained in Ref.~\cite{Marcelis2004}. 
The inset also shows the molecular two-body
bound-state energy as a function of magnetic field,
which shows excellent agreement with the coupled-channels calculation
given in Ref.~\cite{fermi3}.
\label{magnmom}}
\end{figure}

In Fig.~\ref{magnmom} we show the comparison between the
experimental data and the theoretical result from
Eq.~(\ref{eq:mu}), after solving numerically Eq.~(\ref{eq:root}) for 
the molecular bound-state energy and substituting this into
Eq.~(\ref{eq:zeta}).
It is important to consider the full energy
dependencies in Eq.~(\ref{eq:self}), to obtain
the excellent agreement with experiment. Due
to the large values of the background scattering length, an
expansion of the selfenergy for small energies
in the spirit of Fermi liquid theory \cite{Bruun2003}, i.e.,
\begin{eqnarray}
\label{eq:Brunn} &\hbar\Sigma^{\rm
2B}\!\left(E\right)\!\!-\!\hbar\Sigma^{\rm 2B}\!\left(0\right) \!\simeq
\eta(B)\sqrt{-mE^+}\nonumber\\
&\times (1-|a_{\rm{bg}}(B)/\hbar|\sqrt{-mE^+})
\end{eqnarray}
works only very close to resonance when $|\Delta B| < 100$ Gauss, but is unable to
deal with the magnetic-field range accessed experimentally.

Our theoretical curve exhibits close to resonance the same
quantitative deviation from the experiment as the 
coupled-channel calculation
given in Ref.~\cite{fermi3}. 
Jochim {\it et al.}~\cite{fermi3} attribute  
the origin of this systematic deviation to 
the fact that in principle the atoms and molecules experience a different optical trapping potential.
We believe, however, that this explanation is unlikely because the deviation between
the theoretical and experimental results occurs close to resonance 
where $Z(B)\ll 1$ and the dressed molecule wavefunction consists almost completely
of atoms in the open channel.
As an alternative explanation we suggest that many-body
effects could be important in order to describe 
the reduction of the molecular magnetic moment.
Such many-body corrections are expected to become important when
the gas parameter $k_F a(B)$
is no longer small, where $k_F$
refers to the Fermi momentum of the gas. For the densities
used in Ref.~\cite{fermi3} it turns out that in the range 
where the deviation occurs we have that $k_F a(B)\geq 0.1$.
In contrast to coupled-channels calculations, many-body corrections
to the molecular magnetic moment are easily incorporated in our 
quantum field theory approach. They are simply determined by Eqs. (\ref{eq:exact0}-\ref{eq:BetheSalpeter}),
and are an important topic of further investigations.

{\it Conclusions.} --- In this paper we have developed a many-body atom-molecule
theory for broad Feshbach resonances where retardation effects
in the atom-molecule coupling cannot be neglected. The theory is shown
to reproduce the two-body physics exactly. 
As an application, we have calculated the magnetic moment of an atomic $^6$Li gas
and the molecular binding energy.
The agreement with experimental data and with coupled-channels calculations
is excellent.
We believe, therefore, that the results obtained in this Letter constitute a
minimal input for every theory that aims at a complete understanding of the BCS-BEC
crossover physics in atomic $^6$Li near the broad Feshbach resonance at about $834$ Gauss.
Work in this latter direction is presently being completed.

We thank Randy Hulet, Servaas Kokkelmans, and
Bout Marcelis for very stimulating discussions. This work is
supported by the Stichting voor Fundamenteel Onderzoek der Materie
(FOM) and the Nederlandse Organisatie voor Wetenschaplijk
Onderzoek (NWO).

\bibliographystyle{apsrev}

\begin{thebibliography}{104}
\bibitem{greiner2003} M. Greiner, C. A. Regal, and D. S. Jin,
                      Nature {\bf 426}, 537 (2003).
\bibitem{fermi1} K. E. Strecker, G. B. Partridge, and R. G. Hulet,
                 Phys. Rev. Lett. {\bf 91}, 080406 (2003).
\bibitem{fermi2} J. Cubizolles, T. Bourdel, S. J. J. M. F. Kokkelmans,
                 G. V. Shlyapnikov, and C. Salomon,
                 Phys. Rev. Lett. {\bf 91}, 240401 (2003).
\bibitem{fermi3} S. Jochim, M. Bartenstein, A. Altmeyer, G. Hendl, C. Chin,
                 J. H. Denschlag, and R. Grimm,
                 Phys. Rev. Lett. {\bf 91}, 240402 (2003).
\bibitem{zwierlein2003} M. W. Zwierlein, C. A. Stan, C. H. Schunck,
                        S. M. F. Raupach, S. Gupta, Z. Hadzibabic, W. Ketterle,
                        Phys. Rev. Lett. {\bf 91}, 250401 (2003).
\bibitem{stwalley1976} W. C. Stwalley, Phys. Rev. Lett. {\bf 37}, 1628 (1976).
\bibitem{tiesinga1993} E. Tiesinga, B. J. Verhaar, and H. T. C. Stoof,
                       Phys. Rev. A {\bf 47}, 4114 (1993).
\bibitem{stoof1996} H. T. C. Stoof, M. Houbiers, C. A. Sackett, and
                    R. G. Hulet, Phys. Rev. Lett. {\bf 76}, 10 (1996).
\bibitem{timmermans2001} E. Timmermans, K. Furuya, P. W. Milonni,
                         and A. K. Kerman,
                         Phys. Lett. A {\bf 285}, 228 (2001).
\bibitem{ohashi2002} Y. Ohashi and A. Griffin,
                     Phys. Rev. Lett. {\bf 89}, 130402 (2002);
                     Phys. Rev. A {\bf 67}, 033603 (2003);
                     Phys. Rev. A {\bf 67}, 063612 (2003).
\bibitem{milstein2002} J. N. Milstein, S. J. J. M. F. Kokkelmans and
                       M. J. Holland, Phys. Rev. A {\bf 66}, 043604 (2002).
\bibitem{grimm2004}   M. Bartenstein, A. Altmeyer, S. Riedl, S. Jochim, C. Chin,
                      J. H. Denschlag, and R. Grimm,
                      Phys. Rev. Lett. {\bf 92}, 120401 (2004).      
\bibitem{ketterle2004}M. W. Zwierlein, C. A. Stan, C. H. Schunck,
                      S. M. F. Raupach, A. J. Kerman, and W. Ketterle,
                      Phys. Rev. Lett. {\bf 92}, 120403 (2004).
\bibitem{thomas2004}  J. Kinast, S. L. Hemmer, M. E. Gehm, A. Turlapov, and J. E. Thomas,
                      Phys. Rev. Lett. {\bf 92},  150402 (2004).		      
\bibitem{salomon2004} T. Bourdel, L. Khaykovich, J. Cubizolles,
                      J. Zhang, F. Chevy, M. Teichmann, L. Tarruell, S. J. J. M. F. Kokkelmans,
		      and C. Salomon,
                      Phys.~Rev.~Lett. {\bf 93}, 050401 (2004).	
\bibitem{hulet2004}   K. E. Strecker, G. B. Partridge, R. I. Kamar, and R. G. Hulet, 
                     in {\it{Proceedings of the 19th International Conference on Atomic Physics}},
		     to be published by the AIP Press.	      	      
\bibitem{duine2003a} R. A. Duine and H. T. C. Stoof,
                     J. Opt. B: Quantum Semiclass. Opt. {\bf 5}, S212 (2003).
\bibitem{duine2003b} R. A. Duine and H. T. C. Stoof, Phys. Rep. {\bf 396}, 115 (2004).
\bibitem{falco2004a} G. M. Falco, R. A. Duine, and H. T. C. Stoof,
                     Phys. Rev. Lett. {\bf 92},  140402 (2004).
\bibitem{falco2004b} G. M. Falco and H. T. C. Stoof,
                     Phys. Rev. Lett. {\bf 92}, 130401 (2004).
\bibitem{Marcelis2004} E. G. M. v. Kempen, B. Marcelis, and S. J. J. M. F. Kokkelmans,
                       cond-mat/0406722.
\bibitem{randy}     M. Houbiers, H. T. C. Stoof, W. I. McAlexander, and R. G. Hulet,
                      Phys. Rev. A {\bf 57}, R1497 (1998).		       
\bibitem{Bruun2003} G. M. Bruun and C. J. Pethick, Phys. Rev. Lett. {\bf 92}, 140404 (2004).
\end{thebibliography}

\end{document}